\documentclass[aps,twocolumn,eqsecnum,showkeys,aps,superscriptaddress,6pt,pre]{revtex4}
\usepackage{graphicx}
\usepackage{dcolumn}
\usepackage{xspace}
\usepackage{color}

\begin{document}

\title{Heat capacity and susceptibility of rare-earth magnetic superconductors}
\author{Preeti Suman Dash$^1$ and Salila Das}
\affiliation{Department of Physics, Berhampur University,
Berhampur 760007, Odisha, India}
\email{sd.phy@buodisha.edu.in}
\author{Prabir K. Mukherjee}
\affiliation{Department of Physics, Government College of Engineering
and Textile Technology, 12 William Carey Road, Serampore, Hooghly-712201,
India}
\email{pkmuk1966@gmail.com}
\date{\today}

\begin{abstract}
We calculate the specific heat and susceptibility of rare-earth magnetic superconductors in the context of the Ginzburg-Landau theory. The specific heat and susceptibility are calculated both at
 the coexistence phase of antiferromagnetism and superconductivity and normal phase. Theoretical results  are compared with experimental results which agree excellently well.
\end{abstract}
\maketitle
\newpage

\section{Introduction}

 Rare-earth nickel boro-carbide compounds for various R (rare earth) have been studied extensively for the last few years and have shown evidence of superconductivity and magnetism \cite{ RJC, RNC}, distinguishing
 characteristics of these compounds. Rare earth, Rhenium, Dysprosium, Holmium, Erbium, Thulium, Lutetium, and Ytterbium $(R= Dy, Ho, Er, Tm, Lu, and $ Y$)$ compounds have been some
 of the most thoroughly studied superconductors in the past decade. Physical phenomena like the coexistence of superconductivity and long-range magnetic order, the high transition temperature,
 and reenter superconductivity are all essential topics of interest \cite{RJC,RNC,PCC,KHM} on rare earth compounds. The transition temperature of $RNi_2B_2C$  (R = Ho, Dy, Er, and Tm) compounds, with partially filled f-shells
 having a magnetic moment, is relatively low than non-magnetic borocarbides (R = Y, Lu) \cite{KHM,MMH}.  This lower value in $T_C$ points to an interaction between conduction electrons and rare-earth ions.
 These compounds have a large  density of state ( DOS)  values at $E _F$, indicating that Ni states  contribute significantly to the peak \cite{LFM,WEP,RCO}. Many  theoretical and experimental works like superconducting
 order parameters  \cite{LFR,IKY,DSP} , thermal conductivity \cite{KIA,EBR,YMK} , specific heat  \cite{TPE,PRD} , and upper critical field  \cite{NDG,EAS,PSD,TCV}  have been done in these compounds to show the
 coexisting property of superconductivity  and antiferromagnetism. Magnetic susceptibility measurements on the rare earth $DyNi_2B_2C$ show magnetic order at a temperature of 10.5 K  \cite{RJC}. At this temperature, a change 
 in slope followed by a sharp drop at lower temperatures is observed from resistivity measurement implying that this compound has a superconducting phase  \cite{TCV}. At temperatures $T_N = 5.2 K $,  $T_H=5.6K$ , and $T_M=6.0K$ 
 below T = 8k, $ HoNi_2B_2C$ exhibited three distinct anomalies in its specific heat. Neutron data experiment  \cite{NDG,EAS} confirms the transitions  to the Néel structure at $T_N$  below $T_C$,  and the existence of
 antiferromagnetic order in $HoNi_2B_2C$ and  $ErNi_2B_2C$ establishes its exact nature. Below 1.5 K, superconductivity coexists with  antiferromagnetic order in $TmNi_2B_2C$. Thus, $TmNi_2B_2C$ has the most significant 
 difference between $T_C$ and $T_N$ in the entire $RNi_2B_2C$ series, indicating a relatively weak magnetism that does  not destroy but weakens superconductivity on large Fermi surface sheets  \cite{KHM,RMM,LCG}.  At very 
 low temperatures, specific heat implies significant magnetic anisotropy, with ferromagnetic planes weakly coupled antiferromagnetically. A specific heat jump demonstrates superconductivity at the critical temperature $T_C$ for a given field;
 numerous theoretical models based on the two-gap theory have been successfully used to calculate the different properties of these compounds \cite{JJP}.  Our phenomenological formulation used the Ginzburg–Landau model  \cite{PSD,CFJ,DHY}
 to analyze the specific heat  anomalies and susceptibility associated with the onset of superconductivity. Our analysis of the upper critical field using the Ginzburg–Landau model  \cite{PSD} agreed with experimental results, motivating us to use 
the same model to investigate the specific heat and susceptibility of the rare earth magnetic superconductor.
We have presented our model for the specific heat and magnetic susceptibility expressions in Sec. II. The data and results of our investigation  are described in section III, and our conclusions are given in section IV. 
    
\section{ MODEL}

The Landau theory of phase transition is highly general and consequential, and it is used in many different treatments of superconductivity. For example, we know that depending on the independent thermodynamic variable,
 the Gibbs free energy or Helmholtz free energy density is used. Our starting free 
energy will be similar to the free energy of our previous work  \cite{PSD} . So to 
calculate the heat capacity of rare earth magnetic superconductor, the free 
energy density is \cite{PSD}

\begin{widetext}
\begin{eqnarray}
F & = & \int d^3r[F_n + a_1 \left| \psi_a \right| ^2+a_2\left| \psi_b \right| ^2+
\frac 12 b_1 \left| \psi_a \right| ^4+\frac 12b_2\left| \psi_b \right| ^4
+d_1\left| \psi_a \right| ^6
+\alpha (M_a^2+M_b^2)+\frac 12\beta (M_a^4+M_b^4) \nonumber \\
&&+2\delta M_aM_b
+\gamma_1\left| \psi_a \right|^2(M_a^2+M_b^2) 
+\gamma_2\left| \psi_b \right|^2(M_a^2+M_b^2)+2\eta\psi_a^2\psi_b^2-\kappa_1\psi_a^*\psi_b \nonumber \\
&&+\frac {1}{2m_a}\left| (-i\hbar \nabla-\frac {2e\mathbf{A}}{c})\psi_a 
\right| ^2 
+\frac {1}{2m_b}\left| (-i\hbar \nabla-\frac {2e\mathbf{A}}{c})\psi_b 
\right| ^2 \nonumber\\
&&+\kappa_2(i\hbar \nabla-\frac {2e\mathbf{A}}{c})\psi_a^*
(-i\hbar \nabla-\frac {2e\mathbf{A}}{c})\psi_b+\frac {H^2}{8\pi}]
\label{free1}
\end{eqnarray}
\end{widetext}
{\bf Here $\psi_a$ and $\psi_b$ are described as superconducting order parameters 
associated with Ni(3d). $M_a$ and $M_b$ are anti-ferromagnetic order parameter 
of rare earth magnetic superconductor.
The free energy density of the normal phase is described by $F_n$. 
The material parameters $b_1$, 
$b_2$, $d_1$, $\delta$ and $\beta$ are assumed to be positive. 
The coupling constants are defined by $\gamma_1$, $\gamma_2$, $\eta$, 
$\kappa_1$ and $\kappa_2$. For the stability 
of the superconducting state, we assume $\gamma_1>0$ and $\gamma_2>0$.
$e$ is the elementary electron charge. $m_a$ and $m_b$ are defined as 
elementary electron mass. 
As usual, we assume $a_1=a_{01}(T-T_{C1})$, $a_2=a_{02}(T-T_{C2})$, and 
$\alpha=\alpha_0(T-T_{af})$ with $a_{01}>0$, $a_{02}>0$ and $\alpha_0>0$. 
The term $d_1\left| \psi_a \right| ^6$ is the new addition in the free 
energy density (\ref{free1}) compared to the free energy density of our previous work \cite{PSD}.

After the minimization of  Eq. (\ref{free1}) with respect to $\psi_a$, $\psi_b$,
$M_a$ and $M_b$ for uniform system in zero field $H=0$, one can obtain the normal to 
coexistence of antiferromagnetic and 
superconductivity (N-AFS) phase transition with other phase transitions 
\cite{PSD}.

We will now discuss only the N-AFS phase transition since other phase transitions 
were discussed in our previous work \cite{PSD}. The spontaneous magnetization 
($M_{sa}$) in the AFS 
phase ($M_a=-M_b$) can be calculated from Eq. (\ref{free1}). Now the 
minimization of Eq. (\ref{free1}) with respect to $M_a$, $M_b$ and $\psi_b$ 
and substitution of $M_{sa}$ and $\psi_b$ into Eq. (\ref{free1}), we get}

\begin{equation}
F=F_n^*-\frac {a_2^{*2}}{2b_2^*} + a_1^{*} \left| \psi_a \right| ^2
\frac 12 b_1^{*} \left| \psi_a \right| ^4
+ d_1\left| \psi_a \right| ^6
\label{free2}
\end{equation}
where 

$F_n^*=F_n-\frac {\alpha^{2}}{\beta}-\frac {\delta^2}{\beta}+\frac
{\delta \alpha}{\beta}$,

$a_1^{*}=a_1+\frac {2\delta \gamma_1}{\beta}-\frac {2\gamma_1\alpha}{\beta}-\frac {2\eta^*a_2^*}{b_2^*}$,

$b_1^{*}=b_1-\frac {2\gamma_1^2}{\beta}-\frac {2\eta^{*2}}{b_2^*}$,

$a_2^*=a_2+\frac {2\delta \gamma_2}{\beta}-\frac {2\gamma_2\alpha}{\beta}$,

$b_2^*=b_2-\frac {2\gamma_2^2}{\beta}$,

$\eta^*=\eta-\frac {\gamma_1\gamma_2}{\beta}$.

Minimization of Eq. (\ref{free2}) with respect to $\psi_a$, we get 
\begin{equation}
\left|\psi_a\right|^2=R\left[\left(1+\frac{T_C-T}{\Delta T}\right)^{1/2}
-1\right]
\label{wave1}
\end{equation}
where

$R=\frac {b_1^{*}}{6d_1}$,

$\Delta T=\frac {b_1^{*2}}{12a_0^*d_1}$,

$T_C=\frac {T_{C1}a_{01}-\frac {2\delta\gamma_1}{\beta}+\frac {4\delta \eta^*
\gamma_2}{\beta b_2^{*}}-\frac {2\eta^*a_{02}T_{C2}}{b_2^{*}}-
\frac {2\gamma_1\alpha_0T_{af}}{\beta}+\frac {4\eta^*\gamma_2\alpha_0T_{af}}
{\beta b_2^{*}}}{a_0^*}$,

$a_0^*=a_{01}-\frac {2\gamma_1\alpha_0}{\beta}-\frac {2\eta a_{02}}{b_2^{*}}+
\frac {4\eta^* \gamma_2 \alpha_0}{\beta b_2^{*}}$.
 
After the substitution of Eq. (\ref{wave1}) into Eq. (\ref{free2}), the heat capacity can be calculated from $C_P=-T\frac {{\partial}^2 F}{\partial T^2}$ as
\begin{equation}
C_P=\left\{
\begin{array}{ll}
& {\rm }C_0, (T>T_C),
\\
& {\rm }C_0+\frac {3^{1/2}a^{*3/2}_0}{4{d_1}^{1/2}}T(\Delta T+T_{C}-T)^{-1/2}, 
(T<T_C)
\end{array}
.\right.
\label{heat}
\end{equation}
where

$C_0=B+D(T-T_C)+E(T-T_C)^2$ is the background heat capacity.

Here B, D and E are parameters. 

\subsection{Susceptibility}

First we calculate the susceptibility in the AFS phase. When we apply a small
magnetic field $\Delta M$ in the AFS phase, a
small but finite magnetization $\Delta M=M_a+M_b$ 
is produced in the AFS phase.
Then the free energy density (\ref{free1}) for uniform system can be written as

\begin{widetext}
\begin{eqnarray}
F & = & F_n + a_1 \left| \psi_a \right| ^2+a_2\left| \psi_b \right| ^2+
\frac 12 b_1 \left| \psi_a \right| ^4+\frac 12b_2\left| \psi_b \right| ^4
+\alpha (M_a^2+M_b^2)+\frac 12\beta (M_a^4+M_b^4) \nonumber \\
&&+2\delta M_aM_b
+\gamma_1\left| \psi_a \right|^2(M_a^2+M_b^2)
+\gamma_2\left| \psi_b \right|^2(M_a^2+M_b^2)+2\eta\psi_a^2\psi_b^2-\Delta H
\Delta M+\frac {H^2}{8\pi}
\label{free5}
\end{eqnarray}
\end{widetext}

Then by taking $M_a \cong -M_b$, the minimization of free energy density 
(\ref{free5}) with respect to $M_a$ and $M_b$, we obtain
\begin{equation}
(\alpha+\delta)\Delta M+3\beta M_{sa}^2\Delta M+\gamma_1\left| \psi_a \right|^2
\Delta M +\gamma_2\left| \psi_b \right|^2\Delta M-\Delta H=0
\label{diff}
\end{equation}
Then the susceptibility in the AFS phase can be expressed as
\begin{equation}
\chi_{AFS}=\frac {\Delta M}{\Delta H}= \frac {1}{(\alpha+\delta)+3\beta M_{sa}^2+\gamma_1\left| \psi_a \right|^2
+\gamma_2\left| \psi_b \right|^2}
\label{suscep1}
\end{equation}
where 

\begin{equation}
\left|\psi_a\right|^2=\frac {a_1^{**}b_2^*-2\eta^*a_2^*}{4\eta^{*2}-b_1^{**}b_2^*}
\label{wave1a}
\end{equation}

\begin{equation}
\left|\psi_b\right|^2=\frac {a_2^*b_1^{**}-2\eta^*a_1^{**}}{4\eta^{*2}-b_1^{**}b_2^*}
\label{wave2a}
\end{equation}
Here $a_1^{**}$ and $b_1^{**}$ are defined as

$a_1^{**}=a_1+\frac {2\delta \gamma_1}{\beta}-\frac {2\gamma_1\alpha}{\beta}$,

$b_1^{**}=b_1-\frac {2\gamma_1^2}{\beta}$.

After substitution the values of $M_{sa}^2$, $\left| \psi_a \right|^2$ 
and $\left| \psi_b \right|^2$ into Eq. (\ref{suscep1}), the susceptibility
in the AFS phase is given by
\begin{equation}
\chi_{AFS}=\frac {(4\eta^{*2}-b_1^{**}b_2^*)}{G(T-T_{C3})}
\label{suscep2}
\end{equation}
where 

\begin{widetext}
$T_{C3}=\frac {2\gamma_1(2\eta^*a_{02}^*T_{C2}^*-b_2^*a_{01}^*T_{C1}^*)+
2\gamma_2(2\eta^*a_{01}^*T_{C1}^*-b_1^{**}a_{02}^*T_{C2}^*)-(4\delta+2\alpha_0
T_{af})
(4\eta^{*2}-b_1^{**}b_2^*)}{G}$,

$G=2\gamma_1(2\eta^*a_{02}^*-b_2^*a_{01}^*)+2\gamma_2(2\eta^*a_{01}^*-b_1^{**}a_{02}^*)
-2\alpha_0(4\eta^{*2}-b_1^{**}b_2^*)$,

$T_{C1}^*=\frac {a_{01}T_{C1}-\frac {2\gamma_1\alpha_0T_{af}}{\beta}-
\frac {2\delta\gamma_1}{\beta}}{a_{01}^*}$,

$T_{C2}^*=\frac {a_{02}T_{C2}-\frac {2\gamma_2\alpha_0T_{af}}{\beta}-
\frac {2\delta\gamma_2}{\beta}}{a_{02}^*}$,

$a_{01}^*=a_{01}-\frac {2\gamma_1\alpha_0}{\beta}$,

$a_{02}^*=a_{02}-\frac {2\gamma_2\alpha_0}{\beta}$.
\end{widetext}
The susceptibility in the Normal phase can be expressed as
\begin{equation}
\chi_N=\frac {1}{2a_0^*(T-T_C)}
\label{suscep3}
\end{equation}
To calculate the specific heat and susceptibility of various rare-earth compounds, we have solved the equations (\ref{heat}), (\ref{suscep2}) and  (\ref{suscep3} ) with appropriate fitting parameters.

\section{Results and Discussion}

We have  extensively compared   zero-field specific heat on five single crystals $ RNi_2B_2C$  (R= Er, Ho, Dy, Tm) in the range $ 0.1K < T < T_N$  with available experimental results. Figures 1-4 show the specific heat of rare earth compounds against temperature. A prominent peak at $ T_N$ in these compounds indicates a large heat capacity. This large heat capacity below $T_C$ and small positive susceptibility show that antiferromagnetism coexists with superconductivity in these rare earth superconductors.

 Figure 1 shows the variation of specific heat of $ HoNi_2B_2C$ with temperature, and the inset shows susceptibility with temperature, which offers a similar trend as that of the experimental one  \cite {EAS,ADA} . From the graph, we have obtained a peek at $T_N$  and specific heat anomaly below $ T_C$. The significant susceptibility at $T_N$ indicates the transition of the compound from diamagnetism to antiferromagnetism.  The small positive value of susceptibility points towards the coexistence of superconductivity and antiferromagnetism below $T_C$.

In Fig.2, the specific heat of $ ErNi_2B_2C$ shows an anomaly below $T_C$  with a prominent peak at $ T_N$. The pronounced rise in the susceptibility graph with the transition from negative to positive susceptibility below $T_C$ points to the coexistence of superconductivity and antiferromagnetism in Er compounds. Our result shows a similar trend as that of the experiment  \cite{MMR}.

Figure 3 shows the variation of specific heat for the compound of $TmNi_2B_2C$  with temperature and inset graph susceptibility with temperature. This compound forms a similar AF structure to that of $ HoNi_2B_2C$  below $T_N $, with identically arranged moments. Moreover, superconductivity coexists with this AF order below $T_C$ .

 Figure 4 shows the variation of specific heat of  $DyNi_2B_2C$  with temperature. The superconductivity emerges within a well-developed AF order ($T_C < T_N$). The graph shows the coexistence of Superconductivity with AF order below $ T_C$  . Unlike other Ni-based AF superconducting borocarbides, $ DyNi_2B_2C$  shows a prominent hysteresis in the range  $ 1kOe  \leq    T  \leq  10.5 kOe $   \cite{TCV}. In the specific heat calculation, we have neglected the fluctuation of the order parameters. Therefore, fluctuation-driven specific heat calculation is expected to give better results.
The specific heat at the peak along with $T_c$ and $T_N$ for different compounds are given in the tabular form.

\section{Conclusion}

We have calculated specific heat and susceptibility for the rare earth  magnetic superconductors. The variation of specific heat with temperature shows a similar trend as that of available experimental 
observations. The peak in characteristic heat curve at $T_C$ and negative susceptibility in these compounds indicate the coexistence of superconductivity and antiferromagnetism. Our results show good qualitative agreement with 
experimental results. Our specific heat and susceptibility data will open up further empirical findings in these compounds.

\begin{widetext}

\newpage
\bf {Figure captions:}
\begin{figure}[ht]
\includegraphics[height=14.4cm,width=17.4cm,angle=0]{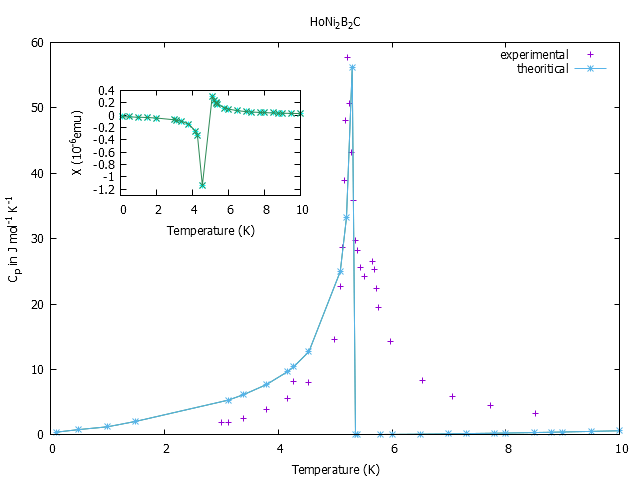}
\caption{ Specific heat as a function of temperature for the Ho compound  along with experimental results  \cite{EAS} for $ 0 \leq T  \leq 10k $ .  Inset: Variation of susceptibility with temperature . The peak at $T_{N}$ corresponds to the antiferromagnetic ordering of the Ho compounds.},
\label{fig1}
\end{figure}
\newpage

\begin{figure}[ht]
\includegraphics[height=14.4cm,width=17.4cm,angle=0]{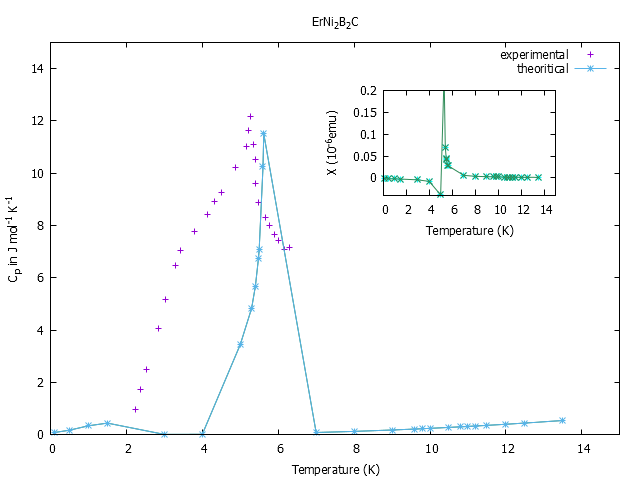}
\caption{ Variation of  specific heat with temperature for $ ErNi_2B_2C$   along with the experimental graph  \cite {MMR}.  The  Inset graph shows the temperature  variation of susceptibility . In both the graphs the peak is observed at  $ T_{N}$ indicating the  presence of antiferromagnetic ordering in  Er compound.},
\label{fig2}
\end{figure}

\newpage
\begin{figure}[ht]
\includegraphics[height=14.4cm,width=17.4cm,angle=0]{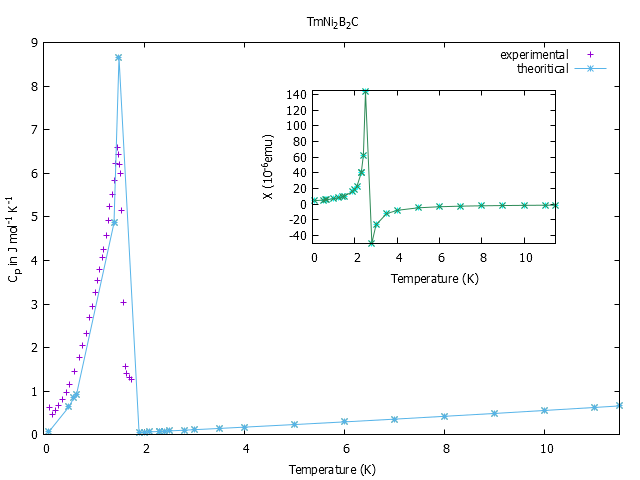}
\caption{The specific heat as a function of temperature for $ TmNi_2B_2C$   along with the experimental graph  \cite{RMM}. The inset graph shows the variation of susceptibility with temperature. The peak at $ T_N$  offers antiferromagnetic ordering of the  Tm compound,}
\label{fig3}
\end{figure}

\newpage
\begin{figure}[ht]
\includegraphics[height=14.0cm,width=16.0cm,angle=0]{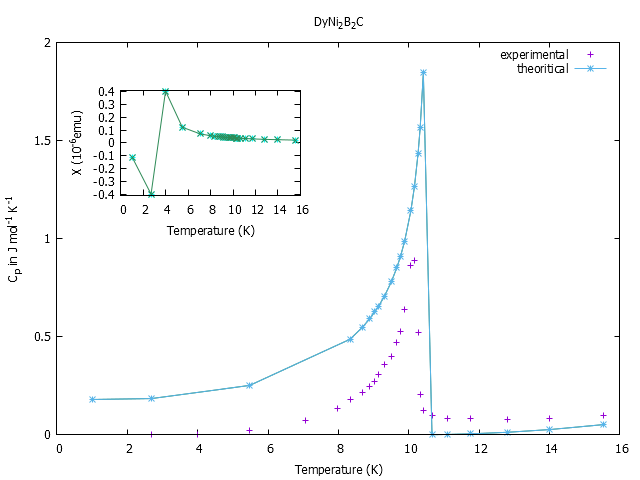}
\caption{ specific heat as a function of temperature for $ DyNi_2B_2C$  along with the experimental graph  \cite{TCV}. The inset graph shows the variation of susceptibility with temperature. The peak at 10.2k  corresponds to  the long-range  antiferromagnetic ordering of the  Tm compound.},
\label{fig4}
\end{figure}

\newpage
\begin{table}
\caption{Values of different specific heat for different compounds}
\begin{ruledtabular}
\begin{tabular}{lllll}
Compounds & $T_C$ in (K) & $T_N$ in (K) & $C_P$ (Theory) in J/mol K & $C_P$ (Experiment) in J/mol K \\
\hline
$HoNi_2B_2C$ & 8.5 & 5.2 & 45.13 & 57.7 \\
$ErNi_2B_2C$ & 11.5 & 6.0 & 11.51 & 12.16\\
$TmNi_2B_2C$ & 11.0 & 1.5 & 8.62 & 6.8\\               
$DyNi_2B_2C$ & 6.5 & 10.5 & 1.84 & 0.9\\
\end{tabular}
\end{ruledtabular}
\end{table}
\end{widetext}

\end{document}